\documentclass[12pt]{article}
\usepackage{epsf}
\def\be{\begin{equation}}
\def\ee{\end{equation}}
\def\bea{\begin{eqnarray}}
\def\eea{\end{eqnarray}}

\begin{document}
\begin{titlepage}
\begin{center}
{\Large \bf William I. Fine Theoretical Physics Institute \\
University of Minnesota \\}
\end{center}
\vspace{0.2in}
\begin{flushright}
FTPI-MINN-19/10 \\
UMN-TH-3819/19 \\
March 2019 \\
\end{flushright}
\vspace{0.3in}
\begin{center}
{\Large \bf Hidden-charm pentaquark formation in antiproton - deuterium collisions
\\}
\vspace{0.2in}
{\bf  M.B. Voloshin  \\ }
William I. Fine Theoretical Physics Institute, University of
Minnesota,\\ Minneapolis, MN 55455, USA \\
School of Physics and Astronomy, University of Minnesota, Minneapolis, MN 55455, USA \\ and \\
Institute of Theoretical and Experimental Physics, Moscow, 117218, Russia
\\[0.2in]

\end{center}

\vspace{0.2in}

\begin{abstract}
The possibility of observing formation of hidden-charm pentaquarks as $s$-channel resonances in antiproton - deuteron collisions is discussed. It is pointed out that
the masses of the reported by LHCb pentaquark resonances in the  $J/\psi \, p$ channel are very close to a special value of the mass at which formation of a pentaquark by antiproton incident on a  deuteron at rest requires exactly the same momentum of the $\bar p$ as needed for the formation in the $s$ channel of the charmonium resonance in $\bar p p$ collisions with the proton being at rest. For this reason the former process can be rather completely described within the notion of the deuteron being a shallow bound state of two nucleons without resorting to models describing its short-distance structure. It is argued that a similar kinematical coincidence can be expected for (yet) hypothetical pentaquark resonances in the $\eta_c \, N$ channel, and that these can be sought for once antiproton - deuterium collisions become available for experimentation. 
  \end{abstract}
\end{titlepage}

The rapidly-growing family of exotic multiquark states containing a heavy quark-antiquark pair, $c \bar c$ or $b \bar b$, has recently been expanded to baryons by the observation~\cite{lhcb1,lhcb2} of the resonances $P_c(4380)$ and $P_c(4450)$ in the hidden-charm pentaquark channel $J/\psi \, p$ produced in the decays $\Lambda_b \to J/\psi \, p \, K^-$.  The new states received explanations in a number of theoretical models, many of which can be found in the recent review~\cite{ghmwzz}. Clearly, a further study of these resonances as well as a search for other baryonic states of similar nature with hidden heavy flavor present a great interest for understanding multi-quark systems. Such studies would certainly be facilitated if other sources of the pentaquark states could be found besides the production in the LHC environment. In particular, it has been pointed out~\cite{wlz,kv,kr} that the observed pentaquark resonances should be produced in the $s$-channel in photoproduction on hydrogen, i.e. in $\gamma +p$ collisions. This conclusion largely does not depend on specific models of the `internal' dynamics of the pentaquark states and is based on the mere fact of the coupling of the resonances to the $J/\psi \, p$ channel and the known interaction of the charmonium state $J/\psi$ with photon. One can readily notice then that in a similar manner the formation of the hidden-charm pentaquarks can be effected by arranging an interaction of a nucleon  with any other state that couples to charmonium. It is the purpose of this paper to point out that a realistic and largely model-independent possibility of producing hidden-charm pentaquarks is offered by the collisions of antiprotons with a deuterium target, which process is possible due to coupling of charmonium states to the $p \bar p$ channel, e.g. $J/\psi \to p \bar p$~\cite{pdg}. Furthermore, this process is possible with other states of charmonium, e.g. it can be used for a search of resonances in the $\eta_c \, n$ channel, which, as will be discussed, has a certain advantage over the $J/\psi$ due to a significantly larger $p \bar p$ decay width of $\eta_c$. Moreover, the discussed formation of hidden-charm pentaquarks is not suppressed by the `softness' of the deutron due to a kinematical coincidence, thus making it plausible that actual experimental searches can be performed by the PANDA experiment~\cite{panda} at FAIR using a deuterium target. 

\begin{figure}[ht]
\begin{center}
 \leavevmode
    \epsfxsize=7cm
    \epsfbox{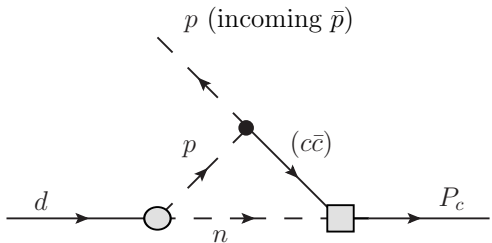}
    \caption{The graph for the hidden-charm pentaquark formation in $\bar p - d$ collision. Dashed lines denote nucleons, the solid lines, as marked, are for the deuteron ($d$), the charmonium state $(c \bar c)$ and the pentaquark $P_c$. }
\end{center}
\end{figure}

The process discussed in this paper is schematically shown in Fig.~1. Clearly, the dominant part of the wave function of the nucleons inside the deuteron, treated as a loose bound state, can be effective only if the kinematical constraints in the graph do not require the relative momentum of the neutron and the proton in the trangle to be large in comparison with the inverse nucleon size. Considering the process in the rest frame of the deutron (which frame coincides with the lab frame in a realistic experiment, e.g. in PANDA), one readily finds that both nucleons in the triangle can be on-shell and simultaneously at rest if the mass $M$ of the pentaquark is related to the mass $m$ of the charmonium state and the nucleon mass $\mu$ as $M=M_0(m)$ with 
\be
M_0^2(m) = 2m^2 + \mu^2~.
\label{m0}
\ee
(The small binding energy $\epsilon = -2.22\,$MeV in the deuteron is obviously neglected in this expression.) In particular, for the charmonium mass of $J/\psi$ and $\eta_c$ the special value of the pentaquark mass is estimated as respectively $M_0(m_{J/\psi}) = 4.48\,$GeV and $M_0(m_{\eta_c}) = 4.33\,$GeV. It can be readily noted that the former of these values is quite close to the measured mass of $P_c(4450)$, while the latter, as will be argued below, is close to the expected mass of a possible pentaquark resonance in the $\eta_c \, N$ channel. With the pentaquark mass within a certain range around the special value $M_0$ the triangle graph of Fig.~1 is dominated by the region of low momentum $|\vec q|$ of each of  the two nucleons, and the wave function of the deuteron can be approximated by that for the free motion 
\be
\phi(\vec q) = {\sqrt{8 \pi \kappa} \over \kappa^2 + \vec q^{\, 2}}~,
\label{ff}
\ee
where $\kappa = \sqrt{ \mu |\epsilon|} \approx 46\,$MeV is the virtual momentum corresponding to the binding energy $\epsilon$ in the deuteron. Clearly, this approximation is valid only as long as the momentum is smaller than the hadronic scale $\Lambda$. Choosing, for an estimate $\Lambda \approx 200\,$MeV, one can find that the condition for the on-shell nucleons in the triangle to have momentum $|\vec q| < \Lambda$ corresponds to the range of the pentaquark resonance masses approximately $4.22 - 5.0$\,GeV for a $J/\psi \, N$ resonance and approximately $4.07 - 4.81$\,GeV for a resonance in the $\eta_c \, N$ channel, with the amplitude decreasing with $M$ away from $M_0$ toward both ends of each of these ranges. 
Outside the mass range where Eq.(\ref{ff}) can be applied any estimate of the amplitude becomes dependent on the model for the short-distance wave function inside the deuteron, and in either case the amplitude is small at those values of the mass. Thus in what follows only the region of $M$ sufficiently close to $M_0$ is considered.

Furthermore, the spin and orbital partial-wave structure of the coupling of $P_c$ to charmonium and a nucleon is not known. Even the spin-orbital structure of the vertex for the decay $J/\psi \to p \bar p$ is known only partially~\cite{babar}. For this reason it would be quite premature to consider various spin and orbital-momentum waves in the vertices in Fig.~1. Such consideration involving an off-shell behavior of the vertices would be necessary for a full calculation.  Here however, for an estimate,  only the absorptive part of the amplitude is evaluated, corresponding to the neutron and the charmonium in the loop being on-shell. Then the kinematical factors associated with the orbital-momentum partial waves in the $P_c \to (c \bar c) \, N$ and the $(c \bar c) \to p \bar p$ vertices are the same as the actual on-shell decay processes, and in fact can be absorbed in the definition of the corresponding vertices, once the rate resulting from the mechanism of Fig.~1 is expressed in terms of the widths $\Gamma[P_c \to (c \bar c) \, N]$ and $\Gamma[(c \bar c) \to p \bar p]$. The spin-dependent factors result in an overall factor of order one and are neglected in the estimates discussed here. Thus in effect all involved particles are considered here as spinless.

The cross section for the process $\bar p + d \to P_c \to anything$ can be evaluated using the well known Breit-Wigner expression (see e.g. in Ref.~\cite{pdg}, Sec. 48) in terms of the branching fraction $Br(P_c \to \bar p \, d)$. At the maximum of the resonance the expression reads as
\be 
\sigma(\bar p + d \to P_c \to anything) = {2 J +1 \over 6} \, {4 \pi \over k^2}  \, Br(P_c \to \bar p \, d)~,
\label{bw}
\ee
where $J$ is the spin of the resonance and $k$ is the c.m. momentum in the decay $P_c \to \bar p \, d$:
\be
k= {\sqrt{(M^2 - \mu^2)(M^2 - 9 \mu^2)} \over 2 M}~.
\label{mk}
\ee
The factor six in the denominator of the overall spin factor in Eq.(\ref{bw}) is the combined total number of spin states for the antiproton and the deuteron. Setting, as discussed above, the spin factor to one (the value of $J$ is not quite known anyway), and considering that for the pentaquark mass in the ballpark of 4.3 - 4.4\,GeV the value of $k$ is approximately 1.6\,GeV, one can find a simple relation for the maximum resonance cross section
\be
\sigma(\bar p + d \to P_c \to anything) \approx Br(P_c \to \bar p \, d) \times 2 \cdot 10^{-27}\, {\rm cm}^2~.
\label{xs}
\ee

In order to evaluate the (absorptive part of) the amplitude given by the graph of Fig.~1, we use nonrelativistic normalization for wave functions of all the particles in the graph and denote as $g$ the vertex for the decay $J/\psi \to p \bar p$ (the solid blob in Fig.~1)  and as $h$ the vertex for $P_c \to (c \bar c) \, n$ (the open rectangle in Fig.~1). The rates of the relevant decays are then expressed, in the chosen normalization,\footnote{In the relevant kinematics some of the particles in the process are in fact relativistic, e.g. the incoming antiproton. This behavior is accounted for by the appropriate extra energy-dependent factors, e.g $2 E_p$ in the equations (\ref{psiw}) and (\ref{pcw}) in comparison with the standard expressions resulting from the relativistic normalization of the wave functions to $2 E$. Also one can notice, as already discussed, that the factors $g$ and $h$ generally depend on kinematic variables which dependence however is the same in the absorptive part of the graph of Fig.~1 and in the on-shell decays.} as
\be
 \Gamma(J/\psi \to p \bar p)= {|g|^2 \over 8 \pi}  \, m \, \sqrt{m^2 - 4 \mu^2}
\label{psiw}
\ee
and
\be
\Gamma(P_c \to J/\psi \, n) = {|h^2| \over 8 \pi \, M^4} (M^2 - m^2+ \mu^2) \, (M^2 +m^2 - \mu^2) \sqrt{[M^2 - (m + \mu)^2] [M^2 - (m - \mu)^2] }~.
\label{pcw}
\ee

The absorptive part of the amplitude of $P_c \to \bar p \, d$, given by the triangle graph of Fig.~1, can be calculated using the wave function of the deuteron in Eq.(\ref{ff}) by means of ordinary (Heitler) perturbation theory in the rest frame of the deuteron, which readily accommodates the nonrelativistic wave function (\ref{ff}) as well as the relativistic motion of some of the involved particles.\footnote{This calculation is similar to those described in Refs.\cite{dv06,bv,mv19} and is equivalent (for the purpose of present treatment) to a calculation in a Lorentz-invariant form~\cite{ghmwz} based on the Weinberg's formula~\cite{Weinberg} for the invariant coupling of a shallow bound state to its constituents.} The expression for the absorptive part $A \equiv -i {\rm Abs} [A(P_c \to \bar p \, d)]$ reads as
\be
A = h \, g \, \int  \pi \, \delta(E_{(c \bar c)} + E_{n} - E_{Pc}) \, \phi(\vec q) \, { d^3 q \over (2 \pi)^3}~,
\label{absa}
\ee
where ${\vec q}$ is the momentum of the neutron, so that the relevant energies in the argument of the $\delta$ function are given as: for the neutron, $E_n=\sqrt{\vec q^{\,2} + \mu^2}$, for the $(c \bar c)$ charmonium, $E_{(c \bar c)}= \sqrt{(\vec p - \vec q)^2 + m^2}$ with $\vec p$ being the momentum (in the rest frame of the deuteron) of the pentaquark (and the same as the momentum of the antiproton). Finally, $E_{P_c}$ is the energy of the produced pentaquark,
\be
E_{P_c} = {M^2 + 3 \mu^2 \over 4 \mu},
\label{epc}
\ee
so that $|\vec p | = \sqrt{E_{P_c}^2 - M^2}$. It can be noticed that the momentum $\vec p$ and the energy $E_{P_c}$ of the pentaquark are rather large: at $M \approx 4.4\,$GeV one finds $|\vec p| \approx 3.8\,$GeV and $E_{P_c} \approx 5.8$\,GeV.

For further calculation it is convenient to choose the $z$ axis of the coordinate system along the momentum $\vec p$ and denote $p_3 = p$, and to consider separately the longitudinal component $q_3$ of the vector $\vec q$ and its transverse part $\vec q_\perp$. The integration in Eq.(\ref{absa}) over the transverse part can be used to remove the $\delta$ function:
\be
\int \delta(E_{(c \bar c)} + E_{n} - E_{Pc}) \, d^2 q_\perp = 2 \pi \, { E_{(c \bar c)} \,  E_{n} \over E_{Pc} }~,
\label{iperp}
\ee
where the value of $q_\perp$ is found from the energy conservation equation
\be
\sqrt{(p-q_3)^2 + q_\perp^2 + m^2} + \sqrt{q_3^2 + q_\perp^2 + \mu^2} - E_{P_c}=0~.
\label{econ}
\ee
Clearly, this equation has a solution for $q_\perp^2$ in the domain, where the expression in the l.h.s is negative at $q_\perp=0$, which requires that $q_3 > Q$ with $Q$ being the solution for the boundary of this domain, 
\be 
\sqrt{(p-Q)^2 + m^2} + \sqrt{Q^2 + \mu^2} - E_{P_c}=0~.
\label{qz}
\ee
The remaining in Eq.(\ref{absa}) integration over $q_3$ runs to the right from $q_3=Q$: 
\be
A= h \, g \, { E_{(c \bar c)} \,  E_{n} \over E_{Pc} } \,\sqrt{\kappa \over 2 \pi} \, \int_{Q}^{\sim \Lambda} \, {dq_3 \over q_\perp^2 + q_3^2}~,
\label{an}
\ee
where the peripheral wave function (\ref{ff}) is used. The integral in the latter formula converges at large $q_3$ and by itself does not require an ultraviolet cutoff $\Lambda$. However,
as previously discussed, the expression for the wave function is applicable only for sufficiently small momentum: $|\vec q | < \Lambda$. Thus the discussed calculation can be justified only if the absolute value of $|Q|$ satisfies the same condition, and this criterion is used in the quoted above estimate of the range of applicability in terms of the pentaquark mass. It can be also noted that the condition $Q=0$ defines the special value of the pentaquark mass in Eq.(\ref{m0}). 

The components of the momentum $\vec q$ are much smaller than all other quantities in Eq.(\ref{econ}). Therefore one can simplify the equation by expanding the l.h.s to the second order in these components resulting in the equation (\ref{econ}) taking the form
\be
q_\perp^2 \, \left (  \sqrt{p^2+m^2}  + \mu \right ) + \left ( q_3^2 - Q^2 \right ) \, \left (  \sqrt{p^2+m^2} + \mu \, {m^2 \over p^2+m^2} \right ) \, - 2 \, (q_3 - Q) \, \mu \, p = 0~,
\label{econ2}
\ee
which relation defines $q_\perp^2$ in the integral in Eq.(\ref{an}) as  a function of $q_3$. Using the solution for the latter equation, one readily finds the integral in the logarithmic approximation:
\be 
A \approx h \, g \, { \sqrt{p^2+m^2}  \over E_{Pc} } \,\sqrt{\kappa \over 2 \pi} \,  { \sqrt{p^2+m^2} + \mu \over 2 \, p } \, \log  \left [ { 2 \, \Lambda \, p \, \mu \over \sqrt{p^2+m^2} \, (\kappa^2 + Q^2) } \right ]~,
\label{anf}
\ee
where, using the approximation of small $Q$, the factors $E_{(c \bar c)}$ and $E_n$ from Eq.(\ref{iperp}) are replaced by their values at $Q=0$, i.e.  respectively by $\sqrt{p^2+m^2}$ and $\mu$.

The formula (\ref{anf})  describes only the absorptive part of the amplitude of the decay $P_c \to \bar p \, d$. However, given the unresolvable at present uncertainty in evaluating the dispersive part, the best that could be done is to make an estimate of the rate including only the absorptive contribution:
\be
\Gamma(P_c \to \bar p + d) \approx {|A|^2 \over 4 \pi}  \, {k \, (M^4 - 9 \mu^4) \over M^3}~,
\label{gpc}
\ee
where $k$ is the c.m. momentum in the decay and is given by Eq.(\ref{mk}).

Using in the expression (\ref{gpc}), the result in Eq.(\ref{anf}), and determining the factors $g$ and $h$ from the Eqs. (\ref{psiw}) and (\ref{pcw}), one thus finds the following somewhat lengthy formula for the branching fraction entering Eq.(\ref{xs}) for the cross section
\bea
&& Br (P_c \to \bar p + d) \approx \nonumber \\
&&Br [P_c \to (c \bar c) + n] \,\Gamma[(c \bar c) \to p \bar p \,] \, {(M^2 -3 \mu^2) \, (M^4 - 10 \, M^2 \mu^2 +16 \, m^2 \mu^2 + 9 \mu^4) \over m \, \sqrt{m^2- 4 \mu^2} \, (M^4 -m^4 +2 \, m^2 \mu^2 - \mu^4 ) (M^2+ 3 \mu^2)} \times \nonumber \\
&& {[4 \, \mu^2 + (M^4 - 10 \, M^2 \mu^2 +16 \, m^2 \mu^2 + 9 \mu^4)^{1/2}]^2 \over   (M^4 - 10 \, M^2 \mu^2 + 9 \mu^4)^{1/2} \, (M^4 + m^4 +\mu^4 - 2 \, M^2 m^2 - 2 \, M^2 \mu^2 - 2\, m^2 \mu^2)^{1/2}} \,  \kappa\, L^2~,
\label{fres}
\eea
with the logarithmic factor $L$ given by
\be
L= \log \left [ { 2 \, \Lambda \, p \, \mu \over \sqrt{p^2+m^2} \, (\kappa^2 + Q^2) } \right ] = \log \left[ {2 \,  \Lambda \mu  (M^4 - 10 \, M^2 \, \mu^2 + 9 \mu^4)^{1/2} \over (M^4 - 10 \, M^2 \mu^2 +16 \, m^2 \mu^2 + 9 \mu^4)^{1/2} \, (\kappa^2 + Q^2) } \right ]~,
\label{lf}
\ee  
and the momentum $p$ is expressed as $p= \sqrt{E_{(c \bar c)}^2 - M^2}$. The  formulas (\ref{fres}) and (\ref{lf}) take a remarkably simple form for the special value of the pentaquark mass given by Eq.(\ref{m0}) and corresponding to $Q=0$:
\be
Br (P_c \to \bar p + d) \approx Br [P_c \to (c \bar c) + n] \, \Gamma[(c \bar c) \to p \bar p \, ] { 8 \,  (m^2-\mu^2) \, \kappa \over 3 \, m \, (m^2 - 4 \mu^2)^{3/2} } \log^2 \left[ {2 \,  \Lambda \mu  \, (m^2-4 \, \mu^2)^{1/2} \over m \, \kappa^2 } \right ]~.
\label{sres}
\ee
Given that the full expressions only moderately depend on $M$ near the special value $M_0$, the latter simplified formula can be used instead for approximate estimates of the cross section in Eq.(\ref{xs}). Proceeding in this way one finds for a pentaquark resonance in the $J/\psi$ channel the estimate
\be
Br (P_c \to \bar p + d) \approx 0.6 \, Br (P_c \to J/\psi + n) \, {\Gamma ( J/\psi \to p \bar p \, ) \over \mu} \approx 10^{-7} \times Br (P_c \to J/\psi + n)~,
\label{psie}
\ee
where the experimental value~\cite{pdg} $\Gamma(J/\psi \to p \bar p \, ) \approx 0.2\,$keV is used. The branching fraction for the pentaquark decay to $J/\psi + N$ is not known. If it is in the ballpark of 10\%, the latter estimate translates into the cross section in Eq.(\ref{xs}) of the order of $10^{-35}$cm$^2$, which looks quite challenging for the expected in the near future experimental capability.

The situation however may be more encouraging if by analogy\footnote{Such analogy is known to work in at least one case of exotic charmonium-like mesonic resonances  $Z_c(4200)$, decaying to $J/\psi + \pi$~\cite{bellezc}, and $Z_c(4100)$ decaying to $\eta_c + \pi$~\cite{lhcbzc}.}  with the pentaquark(s) decaying to $J/\psi + N$ there also exist one or more pentaquarks decaying to $\eta_c + N$, whose mass is by about 100\,MeV lower, and is also close to the corresponding value $M_0(m_{\eta_c})$. Applying the same estimate (\ref{sres}) at the mass $m_{\eta_c}$ one finds
\be
Br (P_c \to \bar p + d) \approx 0.7 \, Br (P_c \to \eta_c + n) \, {\Gamma( \eta_c \to p \bar p \, ) \over \mu} \approx 3 \times 10^{-5} \times Br (P_c \to \eta_c + n)~,
\label{ece}
\ee
where the enhancement mostly comes from a much larger (in absolute terms) $p \bar p$ decay width of  $\eta_c$, $\Gamma(\eta_c \to p \bar p ) \approx 48 \,$keV. The expected cross section in the resonance maximum then amounts to a significant fraction of $10^{-32}$cm$^2$ [assuming $Br (P_c \to \eta_c + n) \sim 10\%$] and , if such pentaquark states do indeed exist, their search appears to be quite feasible at PANDA provided that a deuterium target can be used.

In summary. The formation of hidden-charm pentaquark resonances in $\bar p + d$ collisions is possible with the nucleons moving slowly inside the deuteron due the masses of the pentaquark, charmonium and the nucleon being close to a special kinemtic relation [Eq(\ref{m0})]. The cross section of this process is evaluated by estimating the absorptive part of the amplitude. The numerical value of the expected cross section at the resonance maximum depends on the unknown branching fraction for the decay of pentaquark to charmonium and a nucleon,  $Br [P_c \to (c \bar c) + n]$. If this fraction is of order 10\% the cross section for formation of the resonances in the channel $J/\psi +N$ (in particular, the ones reported by LHCb) is likely quite small, in the ballpark of $10^{-35}$cm$^2$, and may require a significan effort to be observed with the PANDA detector at FAIR. However if similar pentaquark states decaying to $\eta_c + N$ do exist, their formation cross section is estimated to be much larger, due to the much larger $p \bar p$ decay width of $\eta_c$, so that such resonances can be realistically sought for at a luminosity of order $10^{32}$cm$^{-2}$c$^{-1}$ for antiproton collisions with a deuterium target.

This work is supported in part by U.S. Department of Energy Grant No.\ DE-SC0011842.


\begin{thebibliography}{99}
\bibitem{lhcb1} 
  R.~Aaij {\it et al.} [LHCb Collaboration],
  Phys.\ Rev.\ Lett.\  {\bf 115}, 072001 (2015)
  doi:10.1103/PhysRevLett.115.072001
  [arXiv:1507.03414 [hep-ex]].

\bibitem{lhcb2} 
  R.~Aaij {\it et al.} [LHCb Collaboration],
  Phys.\ Rev.\ Lett.\  {\bf 117}, no. 8, 082002 (2016)
  doi:10.1103/PhysRevLett.117.082002
  [arXiv:1604.05708 [hep-ex]].

\bibitem{ghmwzz} 
  F.~K.~Guo, C.~Hanhart, U.~G.~Mei{\ss}ner, Q.~Wang, Q.~Zhao and B.~S.~Zou,
  Rev.\ Mod.\ Phys.\  {\bf 90}, no. 1, 015004 (2018)
  doi:10.1103/RevModPhys.90.015004
  [arXiv:1705.00141 [hep-ph]].
	
\bibitem{wlz} 
  Q.~Wang, X.~H.~Liu and Q.~Zhao,
  Phys.\ Rev.\ D {\bf 92}, 034022 (2015)
  doi:10.1103/PhysRevD.92.034022
  [arXiv:1508.00339 [hep-ph]].
	
\bibitem{kv} 
  V.~Kubarovsky and M.~B.~Voloshin,
  Phys.\ Rev.\ D {\bf 92}, no. 3, 031502 (2015)
  doi:10.1103/PhysRevD.92.031502
  [arXiv:1508.00888 [hep-ph]].

\bibitem{kr} 
  M.~Karliner and J.~L.~Rosner,
  Phys.\ Lett.\ B {\bf 752}, 329 (2016)
  doi:10.1016/j.physletb.2015.11.068
  [arXiv:1508.01496 [hep-ph]].
	
\bibitem{pdg}
M. Tanabashi {\it et al.} [Particle Data Group], 
Phys.\ Rev.\ D {\bf 98}, 030001 (2018).

\bibitem{panda}
  M.~F.~M.~Lutz {\it et al.} [PANDA Collaboration],
  arXiv:0903.3905 [hep-ex].
	
\bibitem{babar} 
  J.~P.~Lees {\it et al.} [BaBar Collaboration],
  Phys.\ Rev.\ D {\bf 87}, no. 9, 092005 (2013)
  doi:10.1103/PhysRevD.87.092005
  [arXiv:1302.0055 [hep-ex]].
	
\bibitem{dv06} 
  S.~Dubynskiy and M.~B.~Voloshin,
  Phys.\ Rev.\ D {\bf 74}, 094017 (2006)
  doi:10.1103/PhysRevD.74.094017
  [hep-ph/0609302].
	
\bibitem{bv} 
  A.~E.~Bondar and M.~B.~Voloshin,
  Phys.\ Rev.\ D {\bf 93}, no. 9, 094008 (2016)
  doi:10.1103/PhysRevD.93.094008
  [arXiv:1603.08436 [hep-ph]].
	
\bibitem{mv19} 
  M.~B.~Voloshin,
  arXiv:1902.01281 [hep-ph].
	
\bibitem{ghmwz} 
  F.~K.~Guo, C.~Hanhart, U.~G.~Mei{\ss}ner, Q.~Wang and Q.~Zhao,
  Phys.\ Lett.\ B {\bf 725}, 127 (2013)
  doi:10.1016/j.physletb.2013.06.053
  [arXiv:1306.3096 [hep-ph]].
	
\bibitem{Weinberg} 
  S.~Weinberg,
  Phys.\ Rev.\  {\bf 137}, B672 (1965).
  doi:10.1103/PhysRev.137.B672
	
\bibitem{bellezc} 
  K.~Chilikin {\it et al.} [Belle Collaboration],
  Phys.\ Rev.\ D {\bf 90}, no. 11, 112009 (2014)
  doi:10.1103/PhysRevD.90.112009
  [arXiv:1408.6457 [hep-ex]].
	
\bibitem{lhcbzc} 
  R.~Aaij {\it et al.} [LHCb Collaboration],
  Eur.\ Phys.\ J.\ C {\bf 78}, no. 12, 1019 (2018)
  doi:10.1140/epjc/s10052-018-6447-z
  [arXiv:1809.07416 [hep-ex]].
	

\end{thebibliography}
\end{document}